\documentclass[11pt,twoside]{article} 
\usepackage{cspm-asp2006}
\usepackage{epsfig,graphicx,natbib,url}  
\usepackage{lscape} 
\begin{document}
\setcounter{page}{271}
\markboth{Heinzel}{Fine Structure of Prominences}
\title{The Fine Structure of Solar Prominences}
\author{Petr Heinzel}
\affil{Astronomical Institute AS, Ond\v{r}ejov, Czech Republic}

\begin{abstract}
Solar prominences and filaments (prominences projected against the solar
disk) exhibit a large variety of fine structures which are well observed
down to the resolution limit of ground-based telescopes. We describe the 
morphological aspects of these fine structures which basically depend on
the type of a prominence (quiescent or active-region). Then we review 
current theoretical scenarios which are aimed at explaining the nature of 
these structures. In particular we discuss in detail the relative roles 
of magnetic pressure and gas pressure (i.e., the value of the plasma-$\beta$), 
as well as the dynamical aspects of the fine structures. Special attention 
is paid to recent numerical simulations which include a complex magnetic 
topology, energy balance (heating and cooling processes), as well as the 
multidimensional radiative transfer. Finally, we also show how new 
ground-based and space observations can reveal various physical aspects
of the fine structures including their prominence-corona transition
regions in relation to the orientation of the magnetic field. 
\end{abstract}

\section{Introduction}

Solar prominences have been observed since the invention of the
spectrohelioscope and as an example one can see well-known H$\alpha$
drawings by \citet{ph-secchi77} (one can be found in the textbook of
\citet{ph-tanhan95}). Although these pioneering observations were
rather simple, they already indicate that prominences consist of many
complicated structures having seemingly chaotic behaviour. Prominence fine
structures have then been frequently observed with better and better
resolution, reaching today a hundred of~km. However, although the fine structure
was known, the prominences have been modeled for decades {\em as a whole}, 
using simplified one-dimensional (1D) slab models. Their magneto-hydrostatic
(MHS) structure was first derived by \citet{ph-ks57} 
(hereafter referred to as KS-model) and
their radiative properties were studied in a 1D-slab approximation by many
authors. Such models reproduced rather well the spectral 
properties of prominences as 
observed with lower resolution. The aim of this review is to discuss various
aspects of current investigations of prominence fine structures. This topic was
thoroughly reviewed by \citet{ph-hv02}, while \citet{ph-engvold04} summarized the 
latest observational results achieved with the highest resolution. Some aspects
of prominence fine structures related to space research were briefly discussed
by \citet{ph-vial06}. For a more general description of the prominence physics
the reader is referred to the monograph of \citet{ph-tanhan95}. 

\section{Morphology of Prominence Fine Structures}

Prominence fine-structure morphology manifests itself rather differently in case of
the limb observations and in case of disk filaments. Moreover, one has to clearly
distinguish between typical quiescent and active-region prominences or filaments.
Generally speaking, a prominence seen on the limb has appeared before or will appear 
later as a filament on the disk. In case of quiescent prominences larger-scale structures
remain fixed while the fine structure changes rather rapidly. However, it seems to be very
difficult to identify the same structures as seen on the limb and on the disk -- this makes 
a lot of confusion and we will discuss this later.

\begin{figure}
  \centering
  \includegraphics[width=0.8\textwidth]{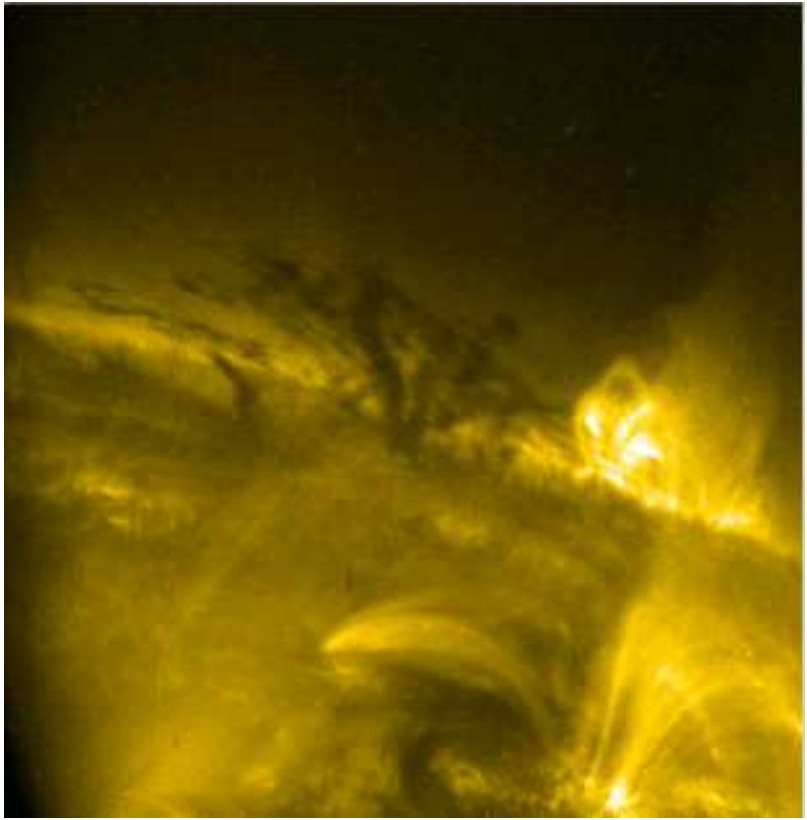}
  \caption[]{\label{ph-fig1}
TRACE image of a limb prominence taken in the 195\,\AA \, line. 
Note the fine structures which appear very dynamic on the respective TRACE movies.
}
\end{figure}

Our information about the fine structure morphology and dynamics comes
from currently available high-resolution ground-based observations
mostly made in the H$\alpha$ line.  With the instruments like the new SST
(Swedish 1-m Solar Telescope) or DOT (Dutch Open Telescope), one can see
fine structures down to the resolution limit (0.15\arcsec\ for SST or
around 100~km). Homogeneous time series are now expected with a
similar spatial resolution from the Solar Optical Telescope (SOT)
onboard the Hinode satellite. A large variety of fine structures and
their dynamics is also seen on TRACE movies, although the spatial
resolution is lower, around 1\arcsec\ (see e.g., TRACE filament movies
on DVD provided by LMSAL). These images are usually taken with a 171 or
195\,\AA\ filter where the hot coronal structures appear
simultaneously with cool ones -- see Fig.~\ref{ph-fig1}. 
Cool prominences or filaments are dark
against the bright background which is due to the absorption of the
background coronal radiation emitted in these lines by the hydrogen
and helium resonance continua (see cartoon in \cite{ph-rutten99}) and
partially due to lack of emissivity of the TRACE lines within a volume
occupied by cool prominence plasmas. At 171 or
195\,\AA, the He\,I and He\,II absorption dominates over the H\,I and it was
shown theoretically that this opacity is quite comparable to that of
the H$\alpha$ line \citep{ph-anhe05}.

\begin{figure}
  \centering
  \includegraphics[width=0.8\textwidth]{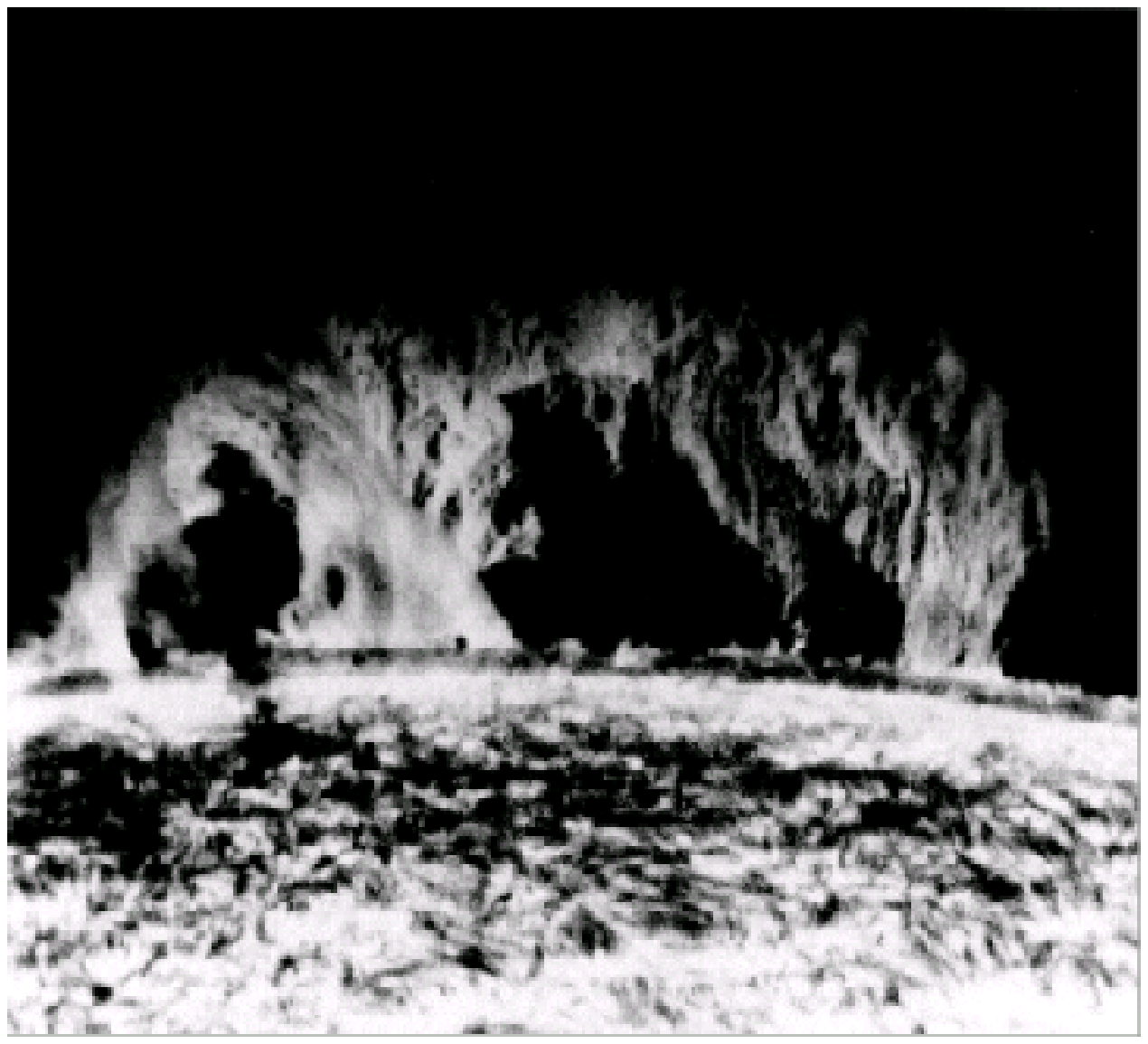}
  \caption[]{\label{ph-fig2}
Big Bear Solar Observatory (BBSO) image of a typical quiescent prominence. 
Quasi-vertical plasma threads are well visible (from \cite{ph-lp05}).
}
\end{figure}

A large quiescent prominence observed at Big Bear Solar Observatory 
(see Fig.~\ref{ph-fig2}) exhibits many
vertically oriented threads of the cool plasma which are a few hundred kilometers wide 
(from \cite{ph-lp05}). 
On a larger scale, one can identify a few vertical plasma sheets which, using a lower
spatial resolution, would appear as more-or-less homogeneous slabs with the thickness typically
smaller compared to the other two spatial dimensions. Such a kind of low-resolution images led 
modelers to use a vertical 1D slab approximation to a real prominence 
geometry which completely neglected the fine structure. In case of MHS models
we already mentioned the classical KS solution. To handle the non-LTE
radiative transfer, 1D slab models were extensively used starting from \citet{ph-polan71},
\citet{ph-yakzel75} and others. We will not detail these models here but rather
discuss their recent generalizations to fine-structure modeling.

\begin{figure}
  \centering
  \includegraphics[width=0.8\textwidth]{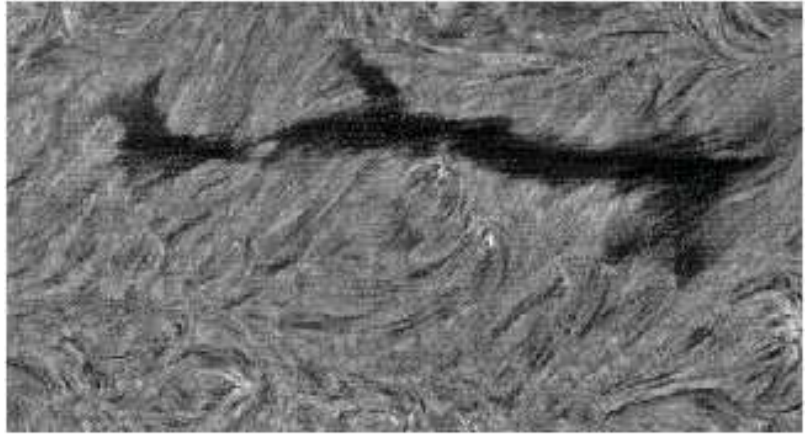}
  \caption[]{\label{ph-fig3}
Dutch Open Telescope (DOT) image of a disk filament. Many dark fibrils are seen
along the filament spine and elsewhere (courtesy of R.J. Rutten).
}
\end{figure}

\begin{figure}
  \centering
  \includegraphics[width=0.7\textwidth]{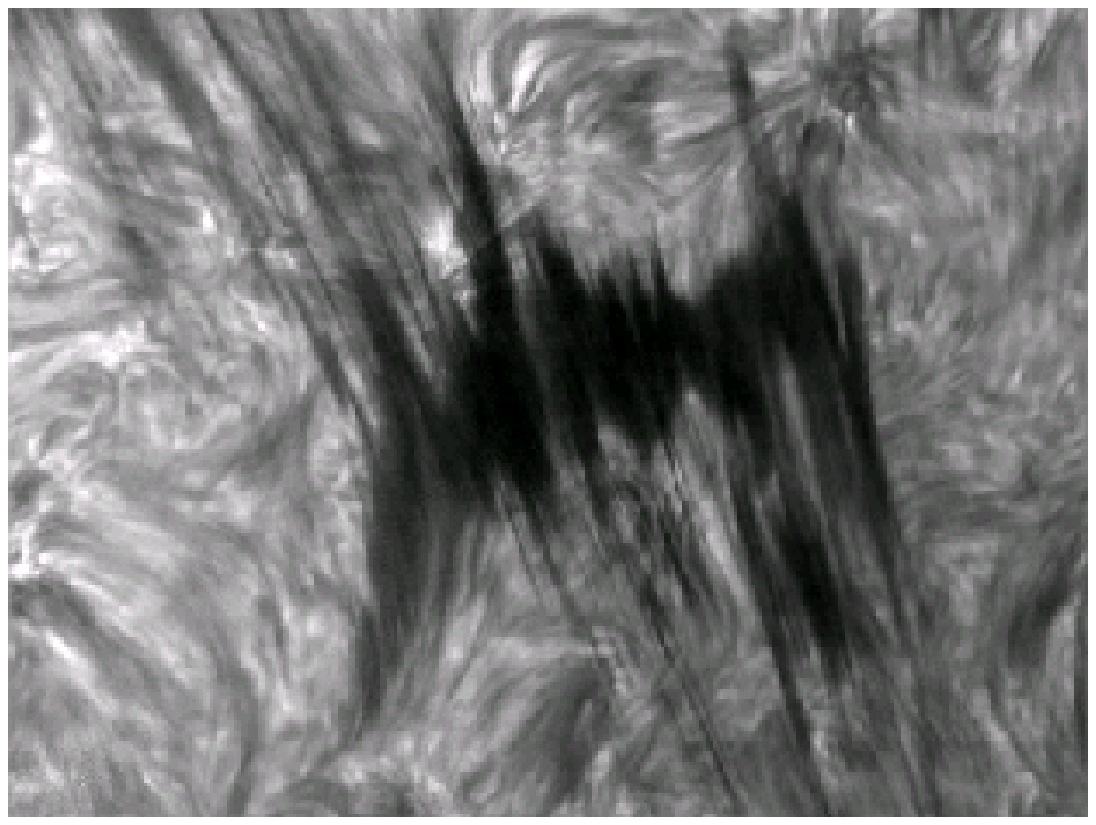}
  \caption[]{\label{ph-fig4}
Swedish 1-m Solar Telescope (SST) image of a disk filament. Here we can see long
dark fibrils which resemble rather thin flux-tubes (courtesy of O.~Engvold). 
}
\end{figure}

On the disk, the high-resolution H$\alpha$ images or movies show
fine-structure fibrils of different lengths, the thinnest visible down to
the resolution limit of SST or DOT. Very thin dark fibrils (we call
them ``fibrils'' to distinguish them from vertical ``threads'' seen on
the limb) visible along the spine of a quiescent filament are rather
short and inclined to the filament axis due to the shear of the
magnetic field lines (Fig.~\ref{ph-fig3}).  On the other hand, much
longer fibrils can be seen within the barbs or connecting various
parts of the filament main body (Fig.~\ref{ph-fig4}). The fact that
densely packed fibrils seen along the spine are rather short compared
to a large-scale magnetic arcade in which the prominence/filament sits
indicates that these fibrils are locations of cool plasma
condensations in a presumably dipped magnetic field.

\section{Dynamics of Prominence Fine Structures}

While the large-scale quiescent prominence structure is rather stable,
the fine structures described above exhibit a strongly dynamical
behaviour. Their shape and brightness change on time scales of minutes
which was already noticed by \citet{ph-engvold76} using the H$\alpha$
prominence observations made at the Dunn VTT at Sacramento Peak.  Today
one can study the fine-structure dynamics and prominence evolution on
prominence/filament movies taken by TRACE in the 195\,\AA \, line or on
high-resolution H$\alpha$ images or movies.  Within the disk
filaments, individual fibrils move sideways with velocities up to
3~km\, s$^{-1}$ which seems to be consistent with limb observations of
\citet{ph-zirkou90,ph-zirkou91}. A qualitatively new observation was
reported by \citet{ph-zirker98} who identified a kind of streaming and
counter-streaming in a filament in both spines and barbs, having flow
velocities around $15 \pm 10$~km\,s$^{-1}$. Dark H$\alpha$ knots were
tracked for distances of 10$^4$ to 10$^5$~km at these speeds -- see long
arrows in Fig.~2 of \citet{ph-zirker98}.

\begin{figure}
  \centering
  \includegraphics[width=0.95\textwidth]{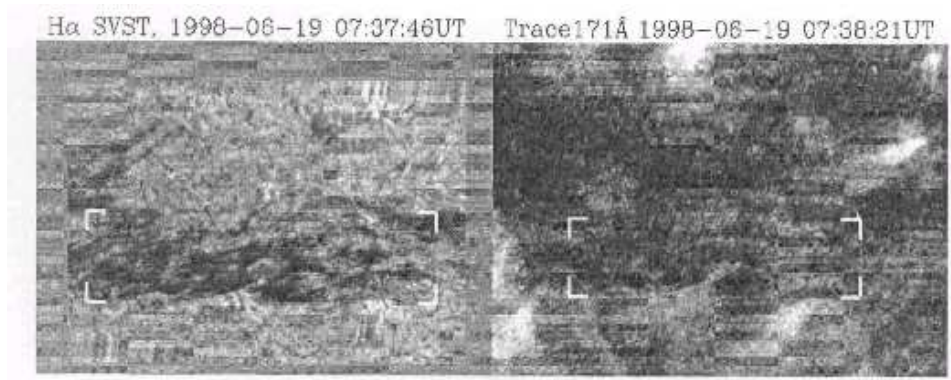}
  \caption[]{\label{ph-fig5}
H$\alpha$ and TRACE fine-structure fibrils. The spatial resolution of TRACE is
lower, but same dark fibrils can be well identified on both images 
(from \cite{ph-slhs04}).
}
\end{figure}

The same counter-streaming was further observed by \citet{ph-lin03} using the high-resolution
SST H$\alpha$ images and Dopplergrams. Both \citet{ph-zirker98} and \citet{ph-lin03}
interpret these motions as plasma flows along the magnetic flux tubes which are not
dipped. We have an evidence that these high-resolution H$\alpha$ images
show fine structures which are well coaligned with dark features visible in the same filament
on TRACE 195\,\AA\, images -- see Fig.~\ref{ph-fig5} from \citet{ph-slhs04}. These authors
have studied the same filament as \citet{ph-lin03}.

\section{Thermodynamic Properties from Spectral Diagnostics}

Spectral diagnostics of prominence fine structures is a difficult task for several reasons.
First, one has to distinguish between optically-thin and optically-thick fine structures. In the former
case several fine-structure elements (FSE) are seen along the line of sight and their radiation
output is thus integrated. On the other hand, in case of thick structures, we mostly see only one
FSE, most probably located closer to the border of a prominence rather than in its central parts.
The situation is actually more complex, because an optically-thick FSE is thick in the line core
but becomes thin in the line wings, where we again see many FSE along the line of sight. This situation
largely complicates the analysis of spectral data. Second reason is that, as already mentioned above, 
FSE are highly dynamical and thus any multi-wavelength observation has to be made simultaneously or at
least quasi-simultaneously (this means within time shorter than the life-time of the FSE) in all
wavelength bands of interest. Another complication arises from the lack of our detailed knowledge
about the internal structure of FSE and their distributions within a prominence -- even 
their elementary geometry is subject of a controversial debate (vertical threads, horizotal
fibrils with flows, blobs etc.). For spectroscopic work, two basic models are usually  considered:
(i) Each FSE has its own prominence-corona transition region
(PCTR), its 3D shape will then depend on the magnetic field threading the FSE
because of principally different thermal conductivity along and across the field lines, (ii) FSE are
more-or-less isothermal, but their temperature increases towards the prominence boundaries and this forms
another kind of PCTR on a larger scale. A combination of (i) and (ii) is also possible. For further details
and references see the review by \citet{ph-vial98} and his recent summary at the SOHO-17 workshop \citep
{ph-vial06}.
The scenario (i) was corroborated  by various authors, while the second one (ii) was suggested e.g.,
by \citet{ph-poleth83} who analyzed the spectral data from SMM/UVSP. Here it is worth mentioning
that the scenario (ii) allows one to use a 1D slab model with a relevant PCTR on each side as a first 
approximation (see \cite{ph-anhe99}), while the effect of a fine-structure PCTR (case (i)) 
has to be modeled by considering
a superposition of several FSE (each of them, however, can be again modeled by a simple 1D slab which
describes, contrary to (ii), only one FSE) -- see \citet{ph-vial89}, \citet{ph-hen89} or \citet
{ph-fon96}.
Finally, a principal difficulty in our understanding of the thermodynamical properties of prominence plasmas
comes from the nature of radiation processes involved. The prominence spectral lines are mostly formed,
namely in cooler parts, by the scattering of the incident radiation coming from the solar surface
(photosphere, chromosphere, or even PCTR, depending on the spectral line under consideration).
This leads to strong departures from LTE and thus the non-LTE theory of radiative transfer must be
applied to prominences and filaments. This is known for a long time, here we can mention pioneering works
by \citet{ph-polan71}, by an HAO group in the seventies (see \citet{ph-hemil78} and references
therein) or by Kiev groups (\cite{ph-yakzel75}, \cite{ph-morozhenko}). For a recent summary of
modern non-LTE techniques see \citet{ph-ha05}. The use of these transfer
methods for 1D slabs is almost routine today and several 1D non-LTE codes do exist: the IAS code
\citep{ph-gl00}, the Ond\v{r}ejov code \citep{ph-hen95}, PANDORA is now also being modified to prominence
slabs \citep{ph-henavr07}. 2D slab models were considered first by \citet{ph-vial82} using the
MAM code of \citet{ph-mam78}, then by \citet{ph-ap94} and \citet{ph-pal95} and finally by
\citet{ph-ha01}. All these codes can be or have been used to model either the prominence or
filament as a whole (note: for prominences one uses slabs standing vertically above the solar surface, 
while in case of filaments the slabs are horizontal, i.e., parallel to the solar surface) or
individual FSE. However,
in case of several FSE, the situation becomes much more complex because of their mutual radiative
interaction. This was treated in simplified ways by \citet{ph-zm89} and by
\citet{ph-hen89}, but in most other cases the radiative interaction was neglected, all FSE 
were considered
the be identical (as 1D slabs) and simply superimposed along the line of sight (\cite{ph-fon96}; \cite{ph-hsvk01}). This interaction could, however, be tackled using todays high-performance parallel machines.

In the above summary we have tried to give the reader an idea how
complex the spectral diagnostics of prominence fine structure
is. However, we still completely neglected the dynamics or temporal
variations of thermodynamic properties. The actual values of
thermodynamic parameters like the kinetic temperature ($T$), gas
pressure ($p$), plasma density ($\rho$), electron density ($n_{\rm
e}$), but also the number of FSE along the line of sight or their
geometrical dimensions, derived from prominence/filament spectra, have
been summarized e.g., by \citet{ph-tanhan95}, \citet{ph-vial98} or
recently, using SOHO/SUMER data, by \citet{ph-pv02}. One approach is
to study in detail the properties at a given prominence location,
where we see -- as discussed above -- one or more FSE along the line
of sight, or to perform a statistical analysis of the raster data to
get the information about a global distribution of such
parameters. Here we can mention the recent work by \citet{ph-wsh07}
(see also \cite{ph-arturo07} on
p.~291\,ff in this volume), where the
line ratio technique (He\,D$_3$ over H$\beta$) was used. The results
based on high-resolution filtrograms taken on VTT at Tenerife do
indicate distinct situations: The case when for example the gas
pressure would be almost constant over the prominence region (see also
\citet{ph-sw05}) is of a particular interest since it indicates an
almost linear decrease of the magnetic field intensity with height and
a statistical homogeneity of the spatial distribution of FSE.

\subsection{Spectral diagnostics with SOHO}

A few months before this Coimbra meeting solar physicists celebrated
10 years of successful observations with the Solar and Heliospheric
Observatory (SOHO), which is a joint ESA/NASA mission (see proceedings
from SOHO 17 Workshop held in Sicily, May 2006 -- ESA SP-617). During
this period, SOHO was frequently used to observe solar prominences and
filaments in UV and EUV spectral bands. In particular SUMER and CDS
spectrometers provided us with a wealth of unique data. The use of SUMER
for prominence physics was reviewed by \citet{ph-pv02}, prominence
spectroscopy with SUMER was also summarized by \citet{ph-hsv06}.
Concerning the prominence fine structure, \citet{ph-wiik99} and
\citet{ph-cirig04} have studied the behaviour of lines formed in PCTR,
while \citet{ph-hsv06} give a summary of their work on the hydrogen
Lyman spectrum. The Lyman lines of hydrogen were also studied by
\citet{ph-cuad03}.  Finally, let us mention that \citet{ph-parenti04}
and \citet{ph-parenti05} have produced an atlas of the prominence EUV
spectrum observed by SOHO/SUMER.

\section{Magnetic Field Determinations}

Magnetic fields in solar prominences were measured since several
decades, using first the Zeeman technique and later on the Hanle
effect. This is summarized in the review by \citet{ph-arturo07} on
p.~291\,ff in this volume. Older
polarimetric observations made with the coronagraph led, in the case
of quiescent prominences, to magnetic fields typically below 20~G.
For latitudes larger than 35$^\circ$ one obtains a mean value of $B$ equal
to 8~G, while at lower latitudes the mean field is around 10~G
\citep{ph-blls94}. \citet{ph-leroy83} analyzed data for a large number
of polar-crown filaments and found $B$ in the range 2 -- 15~G, while
\citet{ph-athay83} arrived at values 6 -- 27~G for a sample of 13
prominences (however, they didn't distinguish between quiescent and
eruptive prominences). Practically all these measurements have
indicated the predominance of horizontal fields which naturally posses
the question why the fine-structure threads appear quasi-vertical and
not aligned along the field lines like the magnetic flux-tubes
\citep{ph-leroy89}.

More recent measurements seem to indicate higher fields, reaching
values of a few tens of Gauss (\cite{ph-arturo07}). However, the respective observations
concern only a rather restricted sample of prominences which may not always be
of a quiescent type or are restricted to low parts of prominences (including
their feets). Higher fields also follow from the analysis of the Stokes $V$ -- signal which
was not used in earlier studies. The actual intensity of the magnetic field is critical
for our understanding of the prominence support and we will discuss this in the
next Section. 

\section{Magnetic Models of Fine Structures}

Since the fine-structure magnetic fields have not yet been detected
(not because of lack of the spatial resolution, but due to still a low
signal-to-noise ratio in polarimetric signals), most of our current
knowledge or ideas come from data-driven modeling. Either one can use
the detailed distribution of surface magnetic fields (like the
SOHO/MDI maps) and perform coronal extrapolations of various degrees
of sophistication (see review by \citet{ph-arturo07} in this volume),
or use indirect spectroscopic methods and modeling to infer the fine
structure of the magnetic field (like that suggested in
\citet{ph-hsvk01} and recently used in \citet{ph-sgha07}).  Here we are
mainly interested in the topology of the field inside and in the surroundings
of FSE (for a recent summary see also \citet{ph-anzer02}).

Two different magnetic configurations are considered in relation to
prominence fine structures.  The first one, also discussed by
\citet{ph-arturo07}, is based on the assumption that
the plasma-$\beta$ (ratio of gas to magnetic pressure) is always very
low, of the order of 10$^{-2}$ or lower. In that case, the plasma
itself has negligible effect on the topology of magnetic fields
related to FSE. If, however, the plasma-$\beta$ is larger, of the
order of 0.1 -- 1.0 or even higher, the weight of the plasma will
produce what we call ``gravity-induced magnetic dips''. Let us discuss
these two situations in more detail (see also \cite{ph-anzer02}).

\subsection{Low-$\beta$ models}

\begin{figure}
  \centering
  \includegraphics[width=0.45\textwidth]{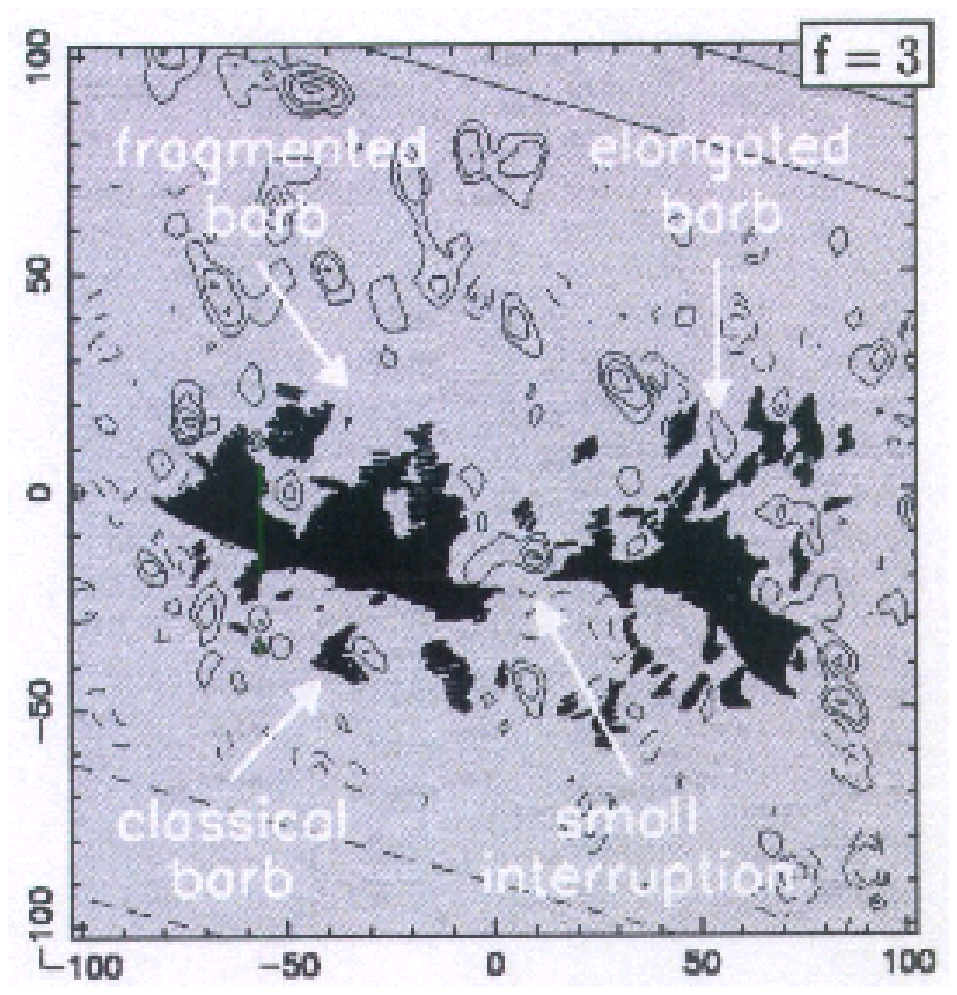}
  \includegraphics[width=0.46\textwidth]{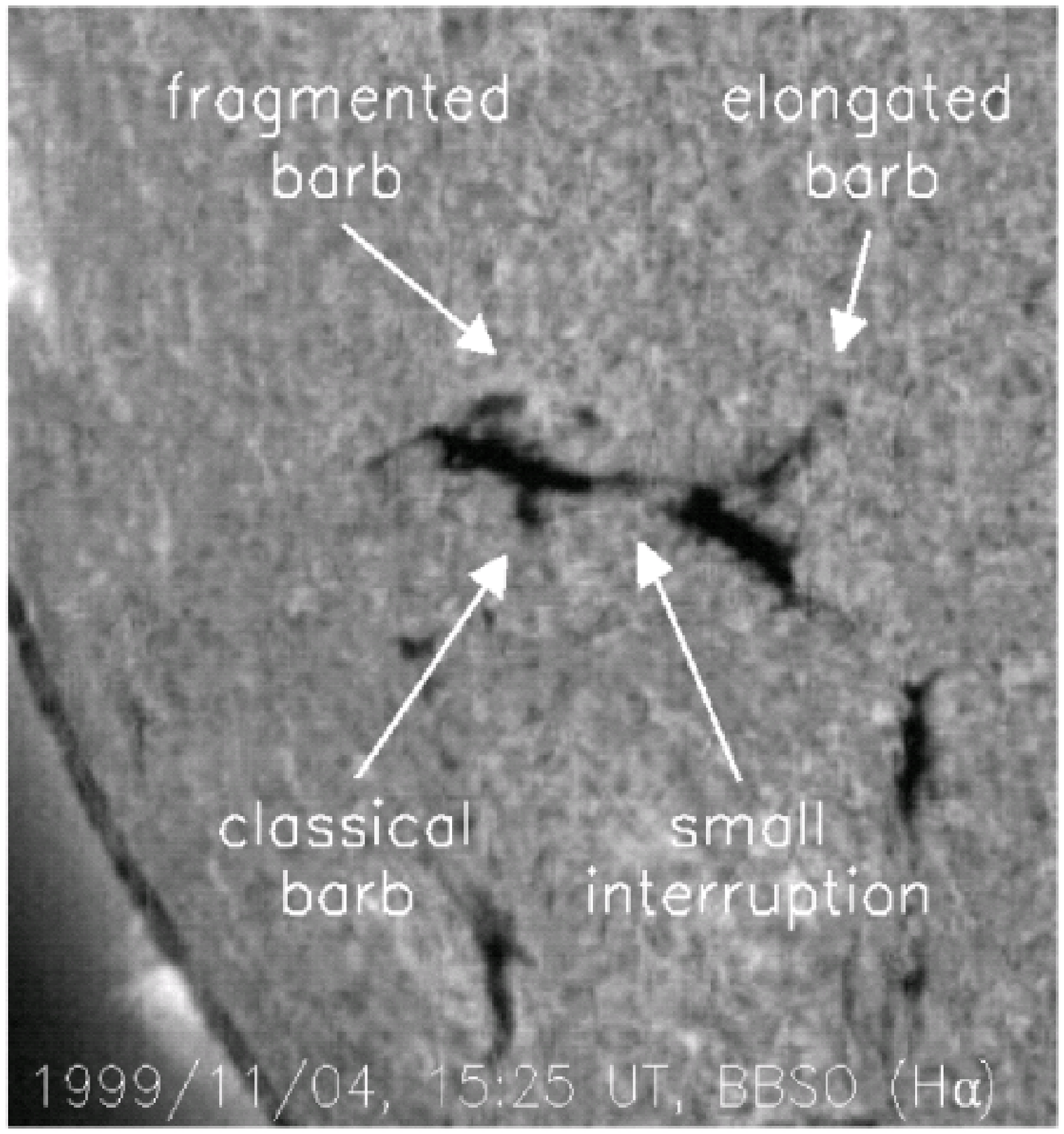}
  \caption[]{\label{ph-fig6}
Computed dark-fibril pattern from linear fff extrapolations (left) and the
corresponding filament observed in the H$\alpha$ line (right). See the text
where this ``blind'' test is described. From \citet{ph-asm00}. 
}
\end{figure}

The topology of the prominence magnetic fields at low $\beta$ is
discussed by \citet{ph-arturo07} on
p.~291\,ff so that we give only a
brief overview here. Today various authors try to model the prominence
magnetic field by linear or non-linear force-free-field (``fff'')
extrapolations of the measured photospheric field. \citet{ph-ad98}
show that the prominence may consist of many rather shallow magnetic
dips which can be filled by the cool material visible e.g., in the
H$\alpha$ line. The presence of such dips was further demonstrated by
\citet{ph-as02} and by \citet{ph-ad03}.  In the latter paper the
authors studied several prominences and found that the extrapolated
field strength is rather low for typically quiescent prominence (less
that 10 G), higher than 10 G for a plage filament and can
reach 30-40 G for active-region prominences.  This clearly shows
how important the knowledge of the prominence type is: when we speak
about the fine structure of quiescent prominences (as in this review),
we cannot use measurements obtained for other types of
structures. Some of the recent measurements which indicate rather
strong fields seem to be related to more active structures. Moreover,
these new measurements were not obtained using the coronagraphs and
thus, because of scattered light higher above the limb, the
measurements are more restricted to lower heights where the
prominences appear brighter and where the polarization signal/noise
ratio is sufficient (L\'{o}pez Ariste -- private communication).  The
usefulness of the filament modeling with linear fff was rather
convincingly demonstrated by \citet{ph-asm00} who performed a kind of
``blind test'': magnetic dips obtained from numerical extrapolation
were marked by black bars schematically indicating the location of the
absorbing H$\alpha$ material -- see Fig.~\ref{ph-fig6}. The shape of
the filament and its various parts were then compared with the true
shape as observed on the disk in H$\alpha$ and the agreement is quite
reasonable.  Non-linear fff models were considered e.g., by
\citet{ph-ball04}.

\begin{figure}
  \centering
  \includegraphics[width=0.7\textwidth]{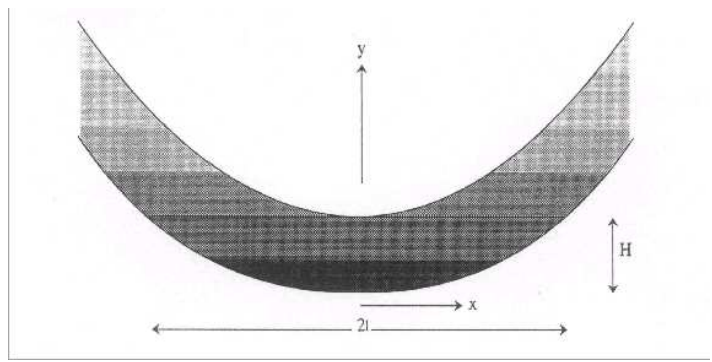}
  \caption[]{\label{ph-fig7}
A schematic model showing how the magnetic dip is filled by the plasma up to
one pressure scale-height $H$. A corresponding horizontal extension of the
mass loading is 2$l$. From \citet{ph-priest}.
}
\end{figure}

For a given curvature of the magnetic dip, \citet{ph-as02} assumed that the dip
is filled by cool plasma up to one  
pressure scale-height $H=kT/g\mu$, where $\mu$ is the mean molecular mass (Fig.~\ref{ph-fig7}). 
For typical prominence conditions $H$ is of the order of 200~km. 
With the computed curvature of the fff dips one can 
draw the bars which indicate the presence of the absorbing material and have the length 2$l$
as indicated in Fig.~\ref{ph-fig7}. However,
the real distribution of the opacity will depend on the actual dip configuration, mass loading in
it and on various plasma parameters including the hydrogen excitation/ionization
conditions. This was not so far considered in the context of fff models.

\subsection{Gravity-induced dip models}

Spectroscopic determinations of the plasma density and its ionization degree, together with spectro-polarimetric
determinations of the magnetic field lead in many cases to rather high values of the plasma-$\beta$. This is for
example the case of a sample of 14 prominences studied by \citet{ph-blls94}, where the plasma-$\beta$
reaches unity in several cases (see also the non-LTE analysis of these prominences by \citet{ph-ha98}).
Using a simple 1D MHS configuration for a vertical fine-structure slab (representing vertical threads as seen in
Fig.~\ref{ph-fig2}), one can easily derive a relation between the plasma-$\beta$ and the angle $\psi$ of the magnetic-field
inclination at the border of the dip (e.g., \cite{ph-ha99}), $\beta \simeq \cot^2 \psi$. 

Construction of equilibrium configurations in which the weight of the
plasma inside the magnetic dip is balanced against the solar gravity
by the Lorentz force is a difficult task. An analytical solution to
this problem, assuming that the whole prominence is represented by a
1D vertical plasma slab hanging in a dipped magnetic field, was
proposed already in 1957 by Kippenhahn and Schl\"{u}ter (so-called
``normal'' polarity model), while \citet{ph-kr74} have suggested a similar model but
with an ``inverse'' polarity. Based on the idea of \citet{ph-pm88} that
the fine-structure threads are in fact vertically-aligned magnetic
dips loaded with mass, \citet{ph-anhe99} and \citet{ph-ha01} started
to model such gravity-induced dips using general 1D or 2D MHS
equilibria, respectively. Such local solutions within the magnetic dip
are sometimes called KS-type models (according to Kippenhahn and
Schl\"{u}ter) because they use the KS-type analytical solution of the
pressure equilibrium (see Section 8). However, they have nothing to do
with the global magnetic topology of the whole prominence so that the
inverse-polarity prominences can be treated in the same
way. A comprehensive description of the gravity-induced dip models is
given in the lecture notes by \citet{ph-ha05} to which the reader
is referred for further details. Here we will only mention one
important result, namely the dependence of the plasma-$\beta$ on
the mass loading $M$ and the field strength $B$, which can be derived
for 1D vertical slab in MHS equilibrium (\cite{ph-anhe07})

\begin{equation}
\beta = \left(\frac{2\pi gM}{B^2}\right)^2.
\end{equation}
Using the results of \citet{ph-ghv}, one can relate this $M$ to optical parameters like
the H$\alpha$ line-center optical thickness or H$\alpha$ integrated intensity.

\begin{figure}
  \centering
  \includegraphics[width=0.7\textwidth]{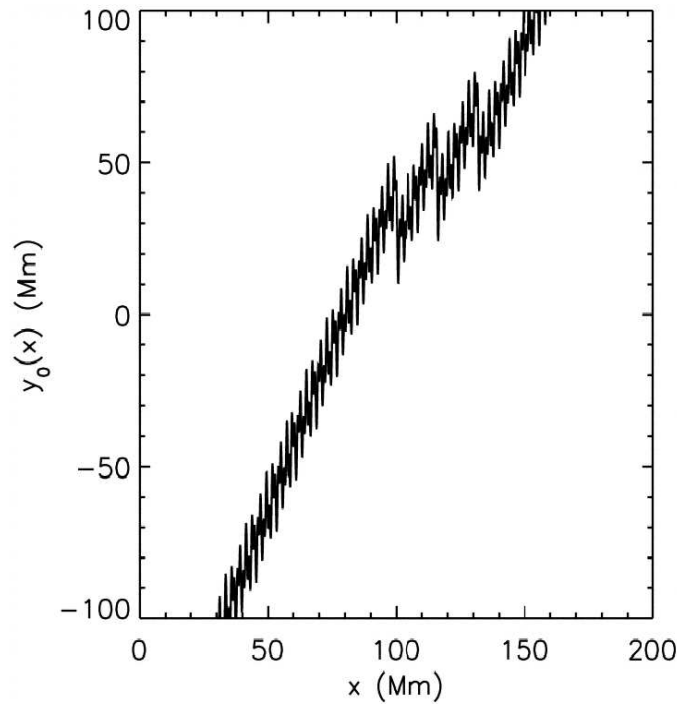}
  \caption[]{\label{ph-fig8}
Gravity-induced magnetic dips as projected onto the disk. Each black bar
corresponds to a vertical 1D dip model. From \citet{ph-lp05}. 
}
\end{figure}

Recently, \citet{ph-lp05} have also considered a series of 1D gravity-induced dips aligned along the 
prominence spine and being in MHS equilibrium of the KS-type (Fig.~\ref{ph-fig8}). However, these authors did not consider
the radiative transfer in these structures and thus could not predict their optical properties like
the H$\alpha$ contrast against the disk. They only draw the black bars as other authors do in
case of low-$\beta$ dips.

\subsection{Other scenarios}

Finally, we should also mention that some authors considered waves
propagating vertically within a flux-tube and aimed at supporting the
cool plasma against the gravity. For example, \citet{ph-peceng00} have
suggested MHD waves. They show an illustratory example (see
their Fig.~1) of such ``vertical flux-tubes'' in a quiescent
prominence with presumably vertical field in which the waves are
propagating.  However, this example rather resembles the situation
discussed by \citet{ph-leroy89}, i.e., quasi-vertical fine-structure
threads which are threaded by horizontal field lines.  There
is also an ongoing debate whether the field in barbs is more inclined or
quasi-vertical (\cite{ph-zirker98}) or made of dips (\cite{ph-as02},
\cite{ph-chae05}). In the former case, for flows which do not
correspond to a free fall, one would need a special support.

\section{Fine-Structure Energetics}

The energetics of quiescent prominences is still an unsolved problem. Although mutually connected, 
we can divide this problem in two aspects: the energetics of central cool cores of the FSE and the
energetics of their PCTRs.

\begin{figure}
  \centering
  \includegraphics[width=0.45\textwidth]{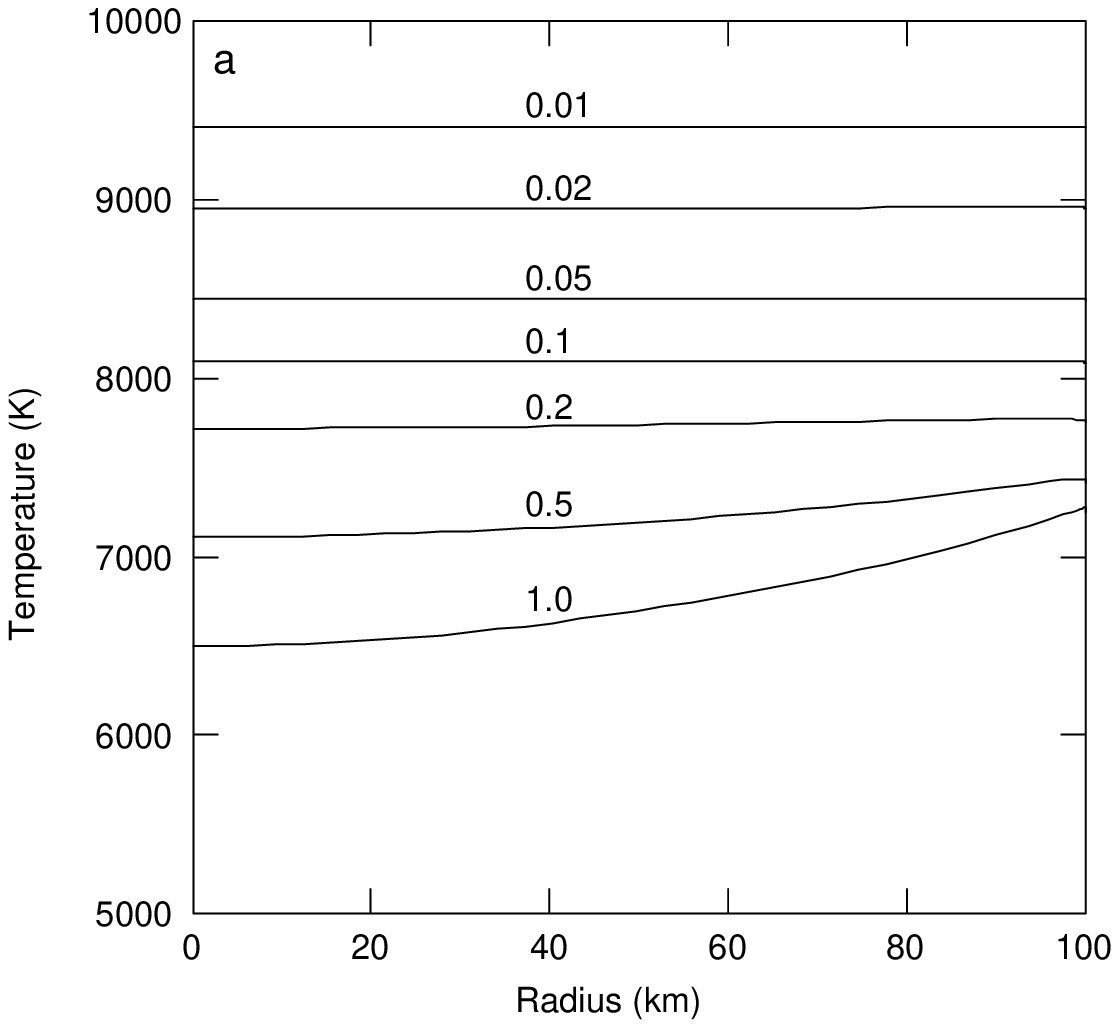}
  \includegraphics[width=0.45\textwidth]{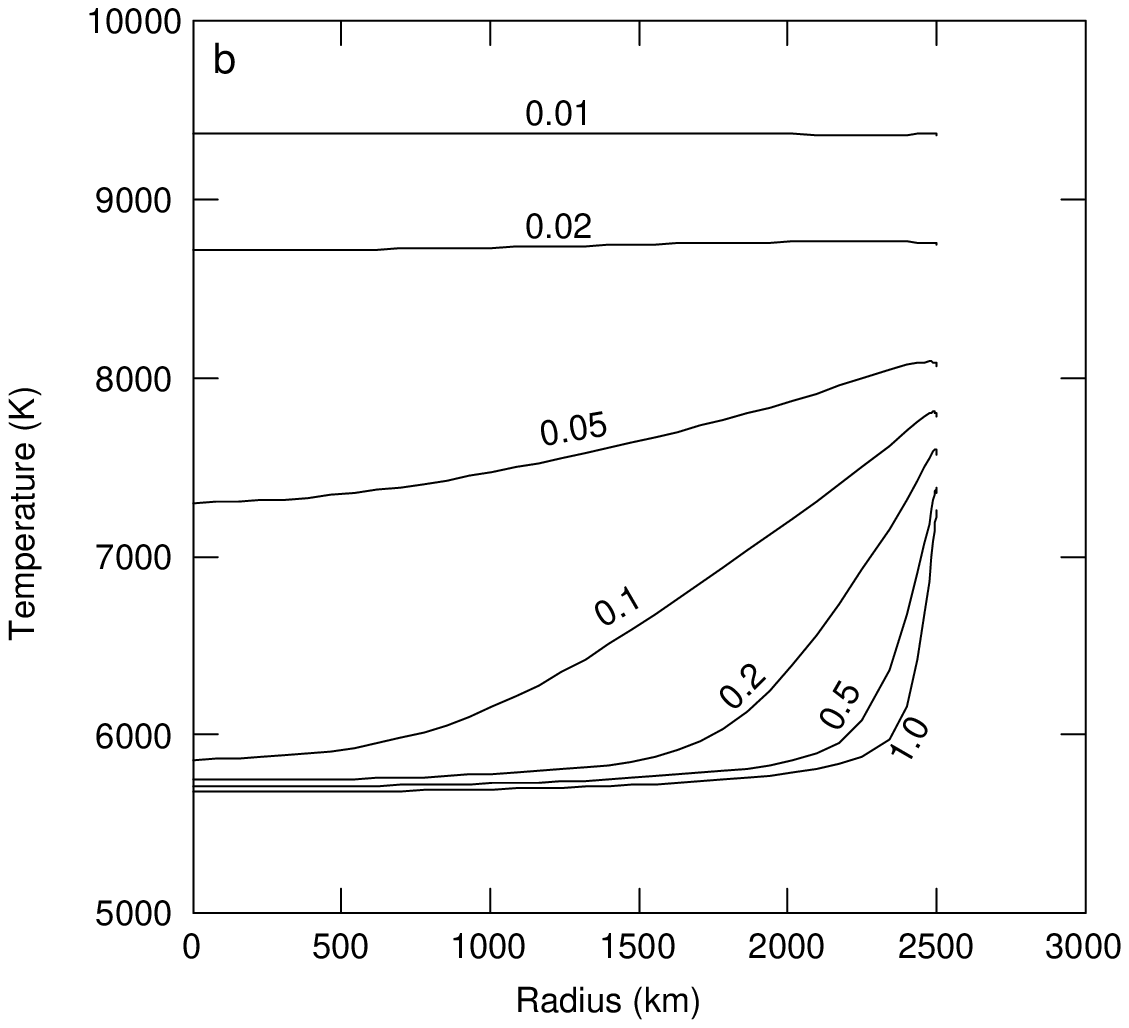}
  \caption[]{\label{ph-fig9}
Temperature variation along the radius of vertical cylindrical threads
in radiative equilibrium, for two diameters and various pressures.
Pressures (in dyn cm$^{-2}$) are indicated as curve labels. Left: 
for cylinders with the width of 200~km; right: for a width of 5000~km. 
From \citet{ph-goutte07}.
}
\end{figure}

The basic question is whether the incident radiation which almost fully determines the radiation
properties of cool cores can also determine the kinetic temperature in these regions. In other words,
is the cool core of FSE in a radiative equilibrium or do we need some extra heating to achieve
the observed values of temperatures which are typically between 6000 to 9000 K. The computation of the radiative equilibrium under the non-LTE
conditions is a rather difficult task, moreover, the result will strongly depend on the incident
radiation fields used as boundary conditions for the solution of the radiation transfer problem.
Radiation-equilibrium models in 1D slabs, for a mixture of hydrogen and helium plasma, 
were first constructed by
\citet{ph-hm76} who arrived at relatively low temperatures, down to 4600 K. By adding some
extra energy input via a heating function, higher temperatures were obtained in accordance with
typical observations. Since the thermal conductivity is not efficient in these central cool parts,
several other mechanisms were considered during the last few decades. Among them, we can mention mainly the
wave dissipation or enthalpy transport.  
However, a surprisingly new result was recently obtained by \citet{ph-goutte07}, who computed a set of
fine-structure models in radiation equilibrium. 1D axially-symmetric cylinders represent vertically standing
threads illuminated by the disk radiation. With currently used incident radiation fields (\cite{ph-goutte04}),
\citet{ph-goutte07} arrived at much higher radiation-equilibrium temperatures as compared to the results of
\citet{ph-hm76} -- we show this in Fig.~\ref{ph-fig9}. Namely thinner cylinders which are of interest for
the fine-structure modeling (plots in Fig.~\ref{ph-fig9}a) reveal temperatures between 7000  and 9500 K for gas
pressures between 0.5 to 0.01 dyn cm$^{-2}$, respectively. This would then mean that no extra heating
is required for cool central parts of fine-strucure threads. In these models only the radiative
losses due to the hydrogen were considered, which nevertheless can be compared with results of \citet{ph-hm76} 
because the helium does not contribute much. However, there is still an open question concerning the
importance of other species like calcium, magnesium or other optically-thin losses (see e.g., discussion
by \citet{ph-anhe99}).

Concerning the energetics of the PCTR, our knowledge is also rather incomplete. At lower temperatures, say
up to 10$^5$~K, the conduction along the magnetic field already plays a role (\cite{ph-fr84}).
Moreover, in analogy to CCTR (\cite{ph-fal}), the ambipolar diffusion was also considered for fine-structure
prominence threads by \citet{ph-fon96}. Quite recently, \citet{ph-ebadi} tested various kinds of 1D models against
the SOHO/SUMER observations of hydrogen Lyman line intensities and found that all models with the ambipolar
diffusion predict unrealistically high integrated intensities of the Lyman $\beta$ line. This can be
related to the relative orientation of the line of sight and the magnetic field as discussed in
\citet{ph-hag05}. At still higher temperatures, most of the work was restricted to the analysis of the differential
emission measure (DEM) -- see e.g., \citet{ph-engvold98}. However, this
analysis is usually based on the so-called coronal approximation and thus the question
arises whether the departures from this approximation, due to radiative excitations, can play any 
important role.
For chromospheric conditions, this issue is discussed in this book by \citet{ph-avrett07}. At temperatures higher
than 30000 K, dynamical models of flux-tubes with prominence condensations were studied using extensive
numerical simulations of the time-dependent loop energetics and plasma dynamics (see \citet
{ph-karp06} and references therein). 

\section{RMHS Simulations}

The equations of magneto-hydrostatic equilibrium originally derived by KS become 
considerably simpler if one uses instead of the Cartesian coordinate $x$ the column-mass 
coordinate $m$, defined by the
relation
\begin{equation}
{\rm d}m = -\rho \, {\rm d}x \, ,
\end{equation}
where $\rho$ is the the plasma density.
$m$ = 0 at one surface of the prominence slab and $m = M$ at
the opposite. Loading of such mass into an initially horizontal magnetic field
leads to a formation of a gravity-induced dip as we will demonstrate in the next
Section. The pressure-balance equation
for such a 1D equilibrium dip configuration is governed by the equation
\begin{equation}
p(m) = 4 p_{\rm c} \frac{m}{M} \left(1 - \frac{m}{M}\right) + p_0 \, ,
\end{equation}
where $p$ is the gas and turbulent pressure and  
$p_0$ is the coronal pressure at the surfaces. This
equation was first derived by \citet{ph-hm76} and used to model prominences
as a whole.
At the slab surface one has the vertical component of the field vector $B_z \equiv B_{z1}$ 
which gives, together with the horizontal component $B_x=$const.
\begin{equation}
M = \frac{B_x B_{z1}}{2\pi g} \, .
\end{equation}
Using this formula, we obtain for $p_{\rm c}$
\begin{equation}
p_{\rm c} = \frac{\pi g^2}{B_x^2} \frac{M^2}{2} = 
\frac{B_{z1}^2}{8\pi} \, .
\end{equation}
The quantity $p_{\rm c}$ can be interpreted in the following way:
at the slab center we have the pressure
\begin{equation}
p_{\rm cen} = p(M/2) = p_{\rm c} + p_0 \, .
\end{equation}
If $p_0$ would be zero, then $p_{\rm cen} = p_{\rm c} = B_{z1}^2/8\pi$,
which is the magnetic pressure. Therefore, in this case                           
the gas pressure at the slab centre
will be equal to the magnetic pressure calculated with $B = B_{z1}$. 
This formulation with the column mass has a great advantage of a simple analytical
integration which is valid for any (i.e., non-constant) temperature and ionization-degree
distribution. 
To get the density $\rho(m)$ we use the state equation with the
mean molecular mass
\begin{equation}
\mu = \frac{1 + 4\alpha}{1 + \alpha + i} m_{\rm H} \, ,
\end{equation}
where $i$ is the ionization degree of hydrogen $i=n_{\rm p}/n_{\rm H}$
($n_{\rm p}$ and $n_{\rm H}$ are the proton and hydrogen densities, respectively), 
$\alpha$ the helium abundance relative to hydrogen and
$m_{\rm H}$ the hydrogen atom mass.
$i$ varies between zero (neutral gas) and unity (fully-ionized
plasma). Inside the prominence and its PCTR, one can consider some schematic variation
of $i$ with depth but for a given prominence model the true 
ionization-degree structure results
from rather complex non-LTE radiative-transfer calculations.

The 1D-slab MHS equilibrium of this type was first used by \citet{ph-hm76}, who combined
it with the full set of non-LTE equations. Their models were aimed at describing the whole
prominence. A similar study was repeated recently by \citet{ph-anhe99} who have considered a
grid of models and studied their energetics. In the latter work, new incident radiation
fields and the partial-redistribution in hydrogen Lyman lines
were used. As a next step, \citet{ph-ha01} have 
generalized these 1D models to two dimensions and developed fully 2D MHS models coupled
to the radiation field. The latter were computed using the 2D radiative-transfer 
technique similar to that
of \citet{ph-ap94}. In this way, all quantities including
the ionization structure were consistently
evaluated in the frame of such a 2D Radiation-MHS (RMHS) approach. However, contrary to previous work,
these 2D models were aimed at representing the vertical fine-structure threads frequently observed
in quiescent prominences. The only drawback of this kind of modeling is that \citet{ph-ha01}
didn't consider the energy-balance problem and, instead, used some kind of empirical temperature
structure. Their PCTR {\em along} the magnetic field lines is much more extended as compared to
that {\em across} the lines, which corresponds to different thermal conductivities. Since these
2D models allow us to look at the fine-structure threads in various directions, we see that
the respective synthetic spectra reflect very well the different structure of PCTR's. Namely
the hydrogen Lyman lines appear quite reversed when we look across the field lines and much
less reversed or even unreversed when looking along the field lines. This quite interesting behaviour
was discovered already in \citet{ph-hsvk01} on basis of SOHO/SUMER spectra.

\begin{figure}
  \centering
  \includegraphics[width=0.45\textwidth]{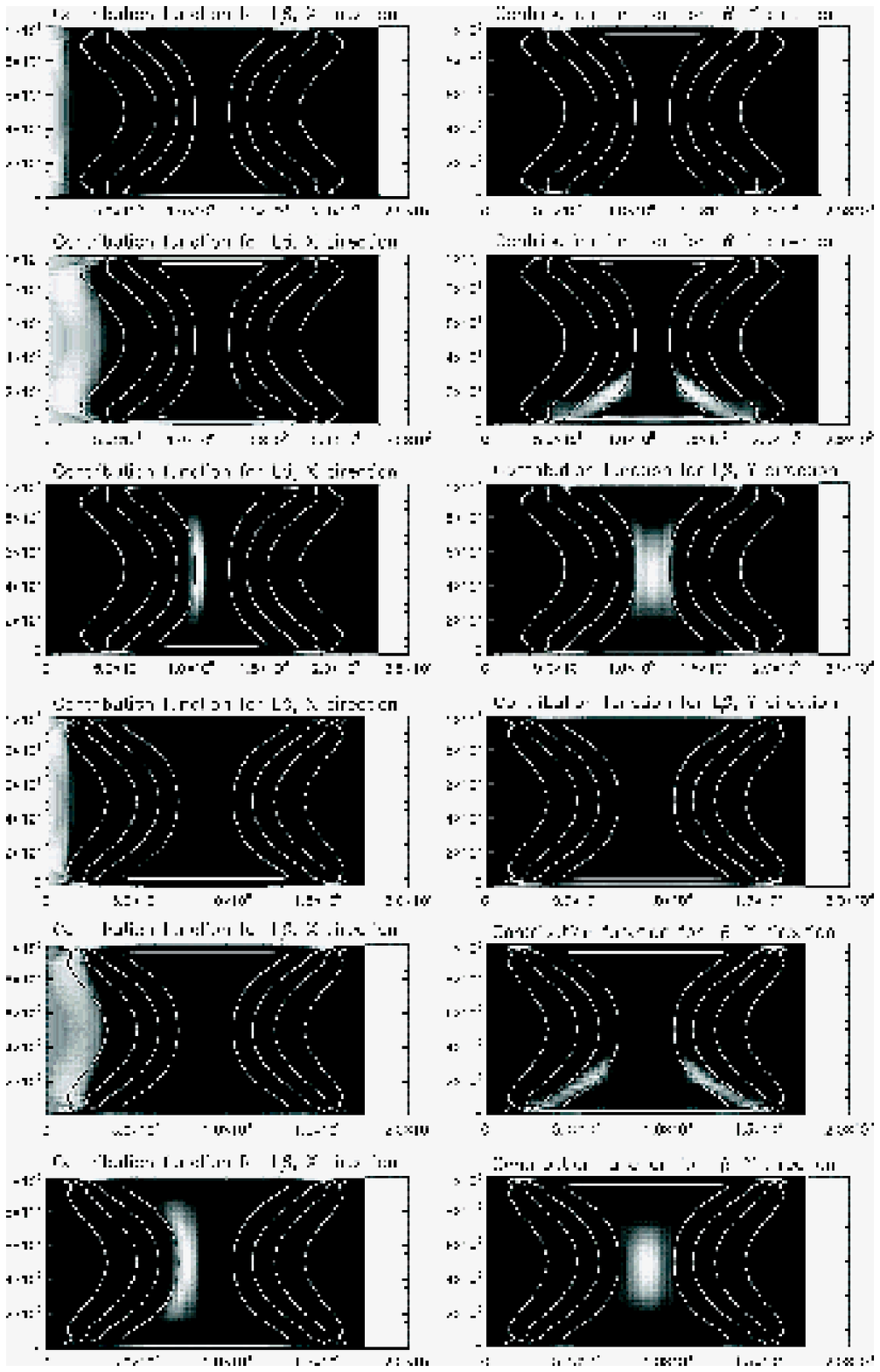}
  \includegraphics[width=0.45\textwidth]{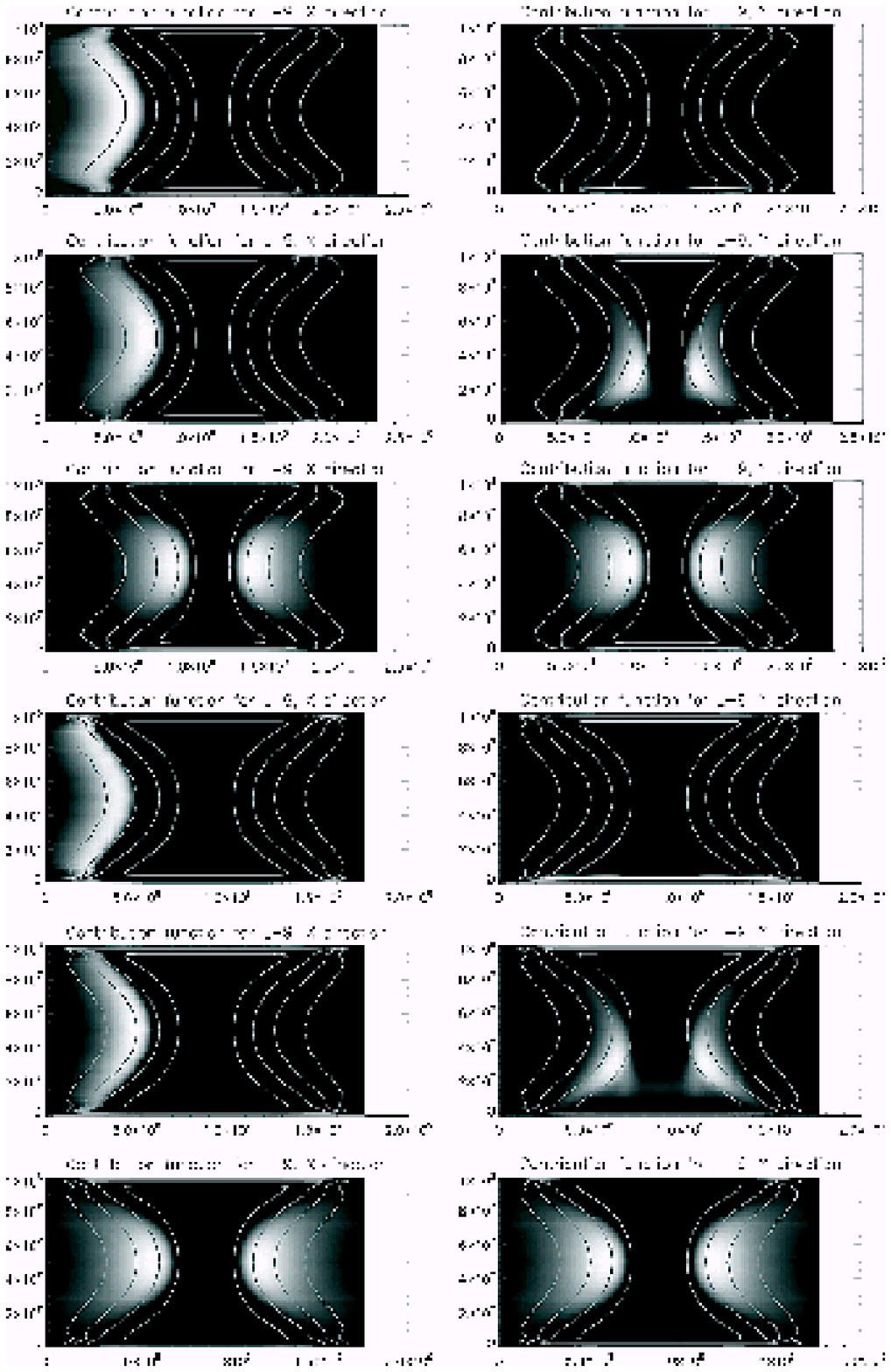}
  \caption[]{\label{ph-fig10}
Contribution functions along and across the field lines showing the formation depths of
the hydrogen Lyman $\beta$ (left two columns) and Lyman 9 (right two columns). Two
different models are displayed (twice three rows). In each part, the first
row corresponds to the line center, the second one to the peak wavelength and the third one to
the wing. Left columns refer to the direction along the field lines ($x$-direction), 
while right columns to the direction across the field ($y$-direction). 
The temperature contours are also drawn showing two quite
different PCTR's. The line core of Lyman $\beta$ is formed at the surface of the 2D
thread (at the highest temperatures), the peaks deeper and in the line wing we can see the central
parts of the structure. The Lyman 9 line is optically thinner and thus the thread more transparent
in all wavelengths allowing the diagnostics of deeper layers. Note the different scales in both
directions, in reality the ``fibril'' projected onto solar disk is much more stretched.
}
\end{figure}

In subsequent papers, \citet{ph-hag05} and \citet{ph-gha07} extended
this 2D modeling of vertical threads to a 12-level plus continuum
hydrogen model atom and studied in detail the formation of hydrogen
Lyman lines and Lyman continuum. In Fig.~\ref{ph-fig10} we show one of
their results, and namely the 2D contribution functions which
illustrate the formation of hydrogen Lyman lines along and across the
field lines, in a 2D vertical thread with PCTR temperatures ranging up
to 50 000 K in this numerical box. This pattern is quite complex but
leads to synthetic line profiles which are in good agreement with
SOHO/SUMER observations (see also the paper by Gun\'{a}r et al.\ on
p.~317\,ff in this volume). The sensitivity of
Lyman-line profiles to orientation of the magnetic field with respect
to the line of sight has been also proven recently by
\citet{ph-sgha07} who studied a kind of ``round-shape filament''
approaching the solar limb and consecutively showing its different
parts above the limb having different orientations of the field lines.

Finally, let us note that these 2D threads which can be anisotropically irradiated represent the
basic ingredient of a more realistic multi-thread modeling. Several such threads distributed in space
will be illuminated by the solar radiation penetrating through this ``forest'' of threads and
the mutual radiative interaction between all those threads can be consistently taken into
account. Another important aspect of this 2D MHS modeling is that such threads or their
clusters can represents prototypes of realistic models used to synthesize the Stokes profiles
and thus study the influence of the fine structure on the polarization signals. 

\section{RMHD Simulations}

\begin{figure}
  \centering
  \includegraphics[width=0.8\textwidth]{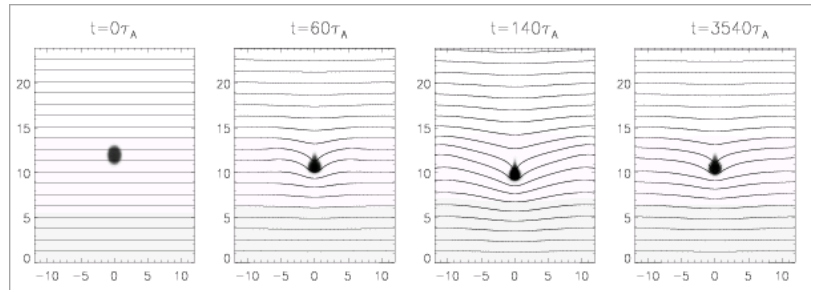}
  \caption[]{\label{ph-fig11}
A sequence of four snapshots of the prominence fine-structure
  evolution according to the MHD system of equations. The simulation is
  carried out in the dimensionless units; lengths are expressed in the
  initial thread half-width $L_{\rm t}$, times in the units of Alfv\'{e}n
  transit time $\tau_{\rm A}=L_{\rm t}/v_{\rm A}$. Recalculated for
  reasonable coronal conditions (coronal temperature 
  $T_{\rm c}=2\times 10^6$K, coronal gravitation length scale 
  $L_{\rm g}=6\times 10^9$~cm) we get $L_{\rm t}\approx 1200$~km, 
  $\tau_{\rm A}\approx 20$~s. Thus the thread characteristic dimension
  is $\approx 2500$~km and the final (relaxed) state in the last frame
  is reached after almost 20 hours.}
\end{figure}

Equation (3) gives the 1D MHS equilibrium  of the KS-type. In order to see
how the cool and dense plasma structure can evolve in the magnetic field
under the action of Lorentz and gravity forces,
\citet{ph-barta07} solved numerically the system of compressible one-fluid 
MHD equations in a 2D vertical
plane, starting with the dense and cool blob (a thread in 3D)
representing a thread of the filament fine structure
surrounded by the gravitationally stratified, constant-temperature hot
corona. The initial magnetic field has only a horizontal component.
At the very beginning the plasma blob starts to fall down, the internal
electric current is induced inside it which generates the restoring Lorenz
force. Due to the inertial mass the blob overshoots the (global)
equilibrium point and an upward-directed Lorentz force prevails over the
gravity turning the movement of the blob up. The system starts to
oscillate but due to the numerical viscous term the oscillations
are damped. After several oscillations the system is practically
relaxed and in the MHS equilibrium. The evolution of the system at
four subsequent times is shown in Fig.~\ref{ph-fig11}. These first simulations
will now be replaced by fully 3D modeling which will also include the 3D
radiative transfer. The latter is necessary to determine the ionization state
of the plasma and its radiation losses.

\section{Comments on Magnetic Dips}

Any analysis of 2D images or movies can lead to misinterpretations just due to the lack
of information on fully 3D structural and dynamical pattern. As an example we will
mention here again the scenario of counter-streaming (\cite{ph-zirker98} or \cite{ph-lin03}).
A close inspection of high-resolution H$\alpha$ movies (see them on CD enclosed to
the Solar Physics Vol. 216) reveals that dark fine-strucure blobs (rather than long threads)
move on relatively short paths, although sometimes in opposite directions. These motions
don't resemble at all flows along long magnetic fluxtubes, but rather a kind of oscillations.
Quite interestingly, when watching the TRACE movies of similar prominences, one clearly sees
sideway oscillations of small blobs which, projected on the disk, would probably resemble the
kind of motions seen on SST movies (projected blobs seen in H$\alpha$
and TRACE 195\,\AA \, are shown in Fig.~\ref{ph-fig5}). However, these blobs visible on TRACE movies form
very frequently the quasi-vertical threads also shown in Fig.~\ref{ph-fig1} and 
Fig.~\ref{ph-fig2} and thus, according
to our discussion above, they should represent plasma condensations hanging in a dipped
magnetic field which seems to oscillate in the corona. This scenario is, however, quite different 
from that of \citet{ph-lin03} (flows along long fluxtubes without any dips).

Another observational support for dips being located along the prominence spine is the
fact that the H$\alpha$ absorbing fibrils are concentrated along the spine and we typically
don't see any significant absorption further away from the spine. This seems to indicate 
that there are no such flows which could be detected along the whole magnetic arcade which
is of course much wider than the width of the spine. Even H$\alpha$ images or movies taken
out of the line center don't show such a pattern.

\begin{figure}
  \centering
  \includegraphics[width=0.7\textwidth]{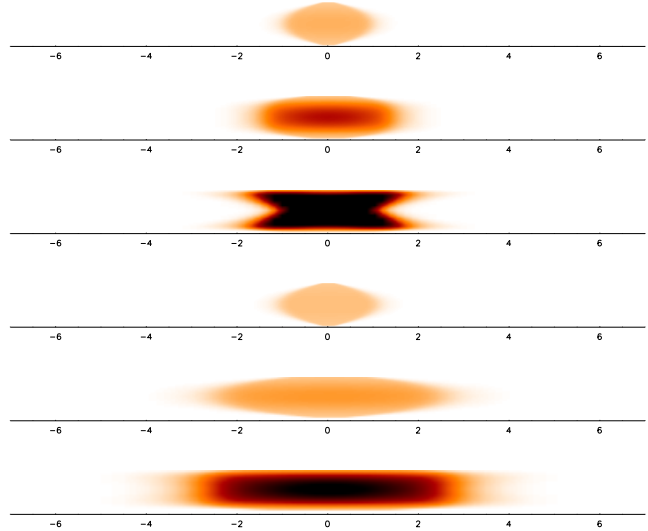}
  \caption[]{\label{ph-fig12}
(2+1)D models of H$\alpha$ fibrils. Six 2D thread models are
shown with different mass loadings and magnetic-field strengths.
From \citet{ph-ha06}.
}
\end{figure}

Actually there should be  no principal controversy between 
low-$\beta$ dip models and gravity-induced dip models. Yet there are no time-dependent 
simulations of how the plasma is condensed in a dipped field and how this may evolve
leading to an enhancement of the plasma density and thus its weight which, eventually, will
modify the magnetic structure of the dip. Existing time-dependent simulations of prominence
condensations assume an initial shape of the magnetic loop and this is not changed
during simulations even when the mass loading at the loop top increases.
Static models of dips which neglect the prominence weight (fff extrapolations) can reasonably
well reproduce the large-scale distribution of dips and this has proven to be a quite
novel approach. On the other hand, static models of gravity-induced dips have been
developed to understand the {\em local} magnetic topology of a dip and its thermodynamic
and radiative properties. These two approaches should not be in conflict, but rather
complementary: provided that the plasma-$\beta$ is large enough, the weight effects 
should be added to initially fff-dips. This is a challenge for future prominence
modeling. As discussed above, today the actual values
of plasma beta are still very uncertain and thus various configurations are possible. An
important diagnostics test would be the quantitative modeling of the H$\alpha$ contrast
of fine-structure fibrils located along the filament spines. This contrast, which certainly
increases with increasing spatial resolution of new telescopes, is due to the absorption
of the background radiation in the H$\alpha$ line (the scattering contribution to the 
source function is typically small). It has to be modeled in 3D fibril geometry which
requires realistic estimate of the mass loading. An example of how this can be handled
was given recently by \citet{ph-ha06}, 
who used some kind of (2+1)D models to demonstrate the theoretical
H$\alpha$ contrast of gravity-induced magnetic dips as projected against the solar disk.
A few examples are shown in Fig.~\ref{ph-fig12}, where the fibril models do exhibit a
significant stretching in the direction of the magnetic field and this resembles real
dark fibrils as seen e.g., in Fig.~\ref{ph-fig3}.

\section{Conclusions}


To conclude this review, let us go again back to Secchi (1877). In his
book {\em Le Soleil} one can read: {\em ``Les protub\'{e}rances se
pr\'{e}sentent sous des aspects si bizarres et si capricieux qu'il est
absolument impossible de les d\'{e}crire avec quelque exactitude.''}
After 130 years, this still seems to be the case.\\

\acknowledgements I am very grateful to my colleagues and
collaborators for many useful discussions which helped me to prepare
this review and to others from SOC and LOC for a nice meeting. In
particular, I am indebted to Uli Anzer and Rob Rutten for reading the
manuscript and suggesting several improvements. This
work was done in the frame of the ESA-PECS project No.\ 98030. I also
acknowledge travel support from the ESMN.



\begin{thebibliography}{}

\bibitem[Anzer(2002)]{ph-anzer02}
Anzer, U. 2002, in Proc. 10th Solar Physics Meeting, ESA SP-506, 389

\bibitem[Anzer \& Heinzel(1999)]{ph-anhe99}
Anzer, U., \& Heinzel, P. 1999, A\&A, 349, 974

\bibitem[Anzer \& Heinzel(2005)]{ph-anhe05}
Anzer, U., \& Heinzel, P. 2005, ApJ, 622, 714 

\bibitem[Anzer \& Heinzel(2007)]{ph-anhe07}
Anzer, U., \& Heinzel, P. 2007, A\&A, in press

\bibitem[Athay et al.(1983)]{ph-athay83}
Athay, R.G., Querfeld, C., Smartt, R., Landi Degl'Innocenti, E., \& Bommier, V. 1983,
Solar Phys., 89, 3

\bibitem[Auer \& Paletou(1994)]{ph-ap94}
Auer, L.H., \& Paletou, F. 1994, A\&A, 285,675

\bibitem[Aulanier \& D\'{e}moulin(1998)]{ph-ad98}
Aulanier, G., \& D\'{e}moulin, P. 1998, A\&A, 329, 1125

\bibitem[Aulanier \& D\'{e}moulin(2003)]{ph-ad03}
Aulanier, G., \& D\'{e}moulin, P. 2003, A\&A, 402, 769

\bibitem[Aulanier \& Schmieder(2002)]{ph-as02}
Aulanier, G., \& Schmieder, B. 2002, A\&A, 386, 1106

\bibitem[Aulanier et al.(2000)]{ph-asm00}
Aulanier, G., Srivastava, N., \& Martin, S. 2000, ApJ, 543, 447

\bibitem[Avrett(2007)]{ph-avrett07} Avrett, E.H., 2007, \newblock in
P.~Heinzel, I. Dorotovi\v{c}, R.~J. Rutten (eds.), The Physics of
Chromospheric Plasmas, ASP Conf.\ Ser.\ 368, 81

\bibitem[Aznar Cuadrado et al.(2003)]{ph-cuad03}
Aznar Cuadrado, R., Andretta, V., Teriaca, L., \& Kucera, T.A. 2003, Mem. S. A. It., 74, 611

\bibitem[B\'{a}rta et al.(2007)]{ph-barta07}
B\'{a}rta, M., Anzer, U., Heinzel, P., \& Karlick\'{y}, M. 2007, in preparation

\bibitem[Bommier et al.(1994)]{ph-blls94}
Bommier, V., Leroy, J.-L., Landi Degl'Innocenti, E., \& Sahal-Br\'{e}chot, S. 1994, 
Solar Phys., 154, 231

\bibitem[Chae et al.(2005)]{ph-chae05}
Chae, J., Moon, Y.J., \& Park, Y.D. 2005, ApJ., 626, 574

\bibitem[Cirigliano et al.(2004)]{ph-cirig04}
Cirigliano, D., Vial, J.-C., \& Rovira, M. 2004, Solar Phys., 223, 95

\bibitem[Engvold(1976)]{ph-engvold76}
Engvold, O. 1976, Solar Phys., 49, 283

\bibitem[Engvold(1998)]{ph-engvold98}
Engvold, O. 1976, in Proc. IAU Coll. 167, New Perspectives on Solar Prominences,
ed. D. Webb, D. Rust \& B. Schmieder, ASP Conf. Ser., 150, 23

\bibitem[Engvold(2004)]{ph-engvold04}
Engvold, O. 2004, in Proc. IAU Symp. No. 223, ed. A.V. Stepanov \& E. Benevolenskaya,
(Cambridge: Cambridge Univ. Press), 187

\bibitem[Fontenla \& Rovira(1984)]{ph-fr84}
Fontenla, J.M., \& Rovira, M. 1985, Solar Phys. 96, 53

\bibitem[Fontenla et al.(1990)]{ph-fal}
Fontenla, J., Avrett, G., Loeser, R. 1990, ApJ, 355, 700 

\bibitem[Fontenla et al.(1996)]{ph-fon96}
Fontenla, J., Rovira, M., Vial, J.-C., \& Gouttebroze, P. 1996, ApJ, 466, 496 

\bibitem[Gouttebroze(2004)]{ph-goutte04}
Gouttebroze, P. 2004, A\&A, 413, 733 

\bibitem[Gouttebroze(2007)]{ph-goutte07}
Gouttebroze, P. 2007, A\&A, in press

\bibitem[Gouttebroze \& Labrosse(2000)]{ph-gl00}
Gouttebroze, P., \& Labrosse, N. 2000, Solar Phys., 196, 349

\bibitem[Gouttebroze et al.(1993)]{ph-ghv}
Gouttebroze, P., Heinzel, P., \& Vial, J.-C. 1993, A\&AS, 99, 513 

\bibitem[Gun\'{a}r et al.(2007)]{ph-gha07}
Gun\'{a}r, S., Heinzel, P., \& Anzer, U. 2007, A\&A, 463, 737 

\bibitem[Heasly \& Mihalas(1976)]{ph-hm76}
Heasley, J.N., \& Mihalas, D. 1978, ApJ, 205, 273                                    

\bibitem[Heasley \& Milkey(1978)]{ph-hemil78}
Heasley, J.N. \& Milkey, R.W. 1978, ApJ, 221, 677                                

\bibitem[Heinzel(1989)]{ph-hen89}
Heinzel, P. 1989, in Proc. IAU Coll. 117, Hvar Obs. Bull. 13, 279

\bibitem[Heinzel(1995)]{ph-hen95}
Heinzel, P. 1995, A\&A, 299, 563

\bibitem[Heinzel \& Anzer(1998)]{ph-ha98}
Heinzel, P., \& Anzer, U. 1998, Solar Phys., 179, 75

\bibitem[Heinzel \& Anzer(1999)]{ph-ha99}
Heinzel, P., \& Anzer, U. 1999, Solar Phys., 184, 103 

\bibitem[Heinzel \& Anzer(2001)]{ph-ha01}
Heinzel, P., \& Anzer, U. 2001, A\&A, 375, 1082

\bibitem[Heinzel \& Anzer(2005)]{ph-ha05}
Heinzel, P., \& Anzer, U. 2005, in Solar Magnetic Phenomena, ed. A. Hanslmeier, A. Veronig \& M.
Messerotti, Astrophys. Space Sci. Lib., 320 (Springer: Dordrecht), 115

\bibitem[Heinzel \& Anzer(2006)]{ph-ha06}
Heinzel, P., \& Anzer, U. 2006, ApJ, 643, L65

\bibitem[Heinzel \& Avrett(2007)]{ph-henavr07}
Heinzel, P. \& Avrett, G. 2007, in preparation

\bibitem[Heinzel \& Vial(1992)]{ph-hv02}
Heinzel, P., \& Vial, J.-C. 1992, Proc. ESA Workshop on Solar Physics and Astrophysics
at Interferometry Resolution, ESA SP-348, 57 

\bibitem[Heinzel et al.(2005)]{ph-hag05}
Heinzel, P., Anzer, U., \& Gun\'{a}r, S. 2005, A\&A, 442, 331

\bibitem[Heinzel et al.(2006)]{ph-hsv06}
Heinzel, P., Schmieder, B., \& Vial, J.-C. 2006, Proc. SOHO 17 -- 10 years of SOHO and Beyond, ESA SP-617 (CD-ROM) 

\bibitem[Heinzel et al.(2001)]{ph-hsvk01}
Heinzel, P., Schmieder, B., Vial, J.-C., \& Kotr\v{c}, P. 2001, A\&A, 370, 281

\bibitem[Karpen et al.(2006)]{ph-karp06}
Karpen, J.T., Antiochos, S.K., \& Klimchuk, J.A. 2006, ApJ, 637, 531

\bibitem[Kippenhahn \& Schl\"{u}ter(1957)]{ph-ks57}
Kippenhahn, R., \& Schl\"{u}ter, A. 1957, Z. Astrophys., 43, 36

\bibitem[Kuperus \& Raadu(1974)]{ph-kr74}
Kuperus, M., \& Raadu, M.A. 1974, A\&A, 31, 189

\bibitem[Leroy(1989)]{ph-leroy89}
Leroy, J.-L. 1989, in Dynamics and Structure of Quiescent Solar Prominences,
ed. E.R. Priest, Astrophys. Space Sci. Lib., 150, 77

\bibitem[Leroy et al.(1983)]{ph-leroy83}
Leroy, J.-L., Sahal-Br\'{e}chot, S., \& Bommier, V. 1983, Solar Phys., 83, 135

\bibitem[Lin et al.(2003)]{ph-lin03}
Lin, Y., Engvold, O., \& Wiik, J.E. 2003, Solar Phys. 216, 109

\bibitem[L\'{o}pez Ariste \& Aulanier(2007)]{ph-arturo07}
L\'{o}pez Ariste, A., \& Aulanier, G. 2007, 
  \newblock in P.~Heinzel, I. Dorotovi\v{c}, R.~J. Rutten (eds.), The
  Physics of Chromospheric Plasmas, ASP Conf.\ Ser.\ 368, 291

\bibitem[Low \& Petrie(2005)]{ph-lp05}
Low, B.C., \& Petrie, G.J.D. 2005, ApJ, 626, 551

\bibitem[Mihalas et al.(1978)]{ph-mam78}
Mihalas, D., Auer, L.H., \& Mihalas, B.W. 1978, ApJ, 220, 1001 

\bibitem[Morozhenko(1984)]{ph-morozhenko}
Morozhenko, N. N. 1984, Spectrophotometric investigations of quiescent solar
prominences (Naukova dumka: Kiev)

\bibitem[Paletou(1995)]{ph-pal95}
Paletou, F. 1995, A\&A, 302, 587

\bibitem[Parenti et al.(2004)]{ph-parenti04}
Parenti, S., Vial, J.-C., \& Lemaire, P. 2004, Solar Phys., 220, 61

\bibitem[Parenti et al.(2005)]{ph-parenti05}
Parenti, S., Vial, J.-C., \& Lemaire, P. 2005, A\&A, 443, 685

\bibitem[Patsourakos \& Vial(2002)]{ph-pv02}
Patsourakos, S., \& Vial, J.-C. 2002, Solar Phys., 208, 253

\bibitem[P\'{e}cseli \& Engvold(2000)]{ph-peceng00}
P\'{e}cseli, H., \& Engvold, O. 2000, Solar Phys., 194, 73

\bibitem[Poland \& Anzer(1971)]{ph-polan71}
Poland, A.I., \& Anzer, U. 1971, Solar Phys., 19, 401

\bibitem[Poland \& Mariska(1988)]{ph-pm88}
Poland, A.I., \& Mariska, J.T. 1988, in Dynamics and Structure of Solar Prominences,
ed. J.L. Ballester \& E.R. Priest (Universit\'{e} des Illes Bal\'{e}ares), 133

\bibitem[Poland \& Tandberg-Hanssen(1983)]{ph-poleth83}
Poland, A.I., \& Tandberg-Hanssen, E. 1983, Solar Phys., 84, 63

\bibitem[Priest(1990)]{ph-priest}
Priest, E.R. 1990, in Proc. IAU Coll. 117, ed. V. Ru\v{z}djak \& E. Tandberg-Hanssen,
Lecture Notes in Physics, 363 (Springer-Verlag: Berlin), 150 

\bibitem[Rutten(1999)]{ph-rutten99}
Rutten, R.J. 1999, in Proc. 3rd Advances in Solar Physics Euroconference: Magnetic Fields
and Oscillations, ed. B. Schmieder, A. Hofmann \& J. Staude, ASP Conf. Ser., 184 (ASP: San
Francisco), 181

\bibitem[Secchi(1877)]{ph-secchi77}
Secchi, A. 1877, Le Soleil (Gauthier-Villars: Paris)

\bibitem[Schmieder et al.(2007)]{ph-sgha07}
Schmieder, B., Gun\'{a}r, S., Heinzel, P., \& Anzer, U. 2007, Solar Phys., in press

\bibitem[Schmieder et al.(2004)]{ph-slhs04}
Schmieder, B., Lin, Y., Heinzel, P., \& Schwartz, P. 2004, Solar Phys., 221, 297

\bibitem[Stellmacher \& Wiehr(2005)]{ph-sw05}
Stellmacher, G., \& Wiehr, E. 2005, A\&A, 431, 1059

\bibitem[Tandberg-Hanssen(1995)]{ph-tanhan95}
Tandberg-Hanssen, E. 1995, The Nature of Solar Prominences (Kluwer: Dordrecht)

\bibitem[van Ballegooijen(2004)]{ph-ball04}
van Ballegooijen, A.A. 2004, ApJ, 612, 519

\bibitem[Vial(1982)]{ph-vial82}
Vial, J.-C. 1982, ApJ, 254, 780

\bibitem[Vial(1998)]{ph-vial98}
Vial, J.-C. 1998, in Proc. IAU Coll. 167, New Perspectives on Solar Prominences,
ed. D. Webb, D. Rust, \& B. Schmieder, ASP Conf. Ser., 150, 175

\bibitem[Vial(2006)]{ph-vial06}
Vial, J.-C. 2006, Proc. SOHO 17 -- 10 years of SOHO and Beyond, ESA SP-617 (CD-ROM)

\bibitem[Vial et al.(2007)]{ph-ebadi}
Vial, J.-C., Ebadi, H., \& Ajabshirizadeh, A. 2007, Solar Phys., in press 

\bibitem[Vial et al.(1989)]{ph-vial89}
Vial, J.-C., Rovira, M., Fontenla, J., \& Gouttebroze, P. 1989, in Proc. IAU Coll.
117, Hvar Obs. Bull. 13, 347

\bibitem[Wiehr et al.(2007)]{ph-wsh07}
Wiehr, E., Stellmacher, G., \& Hirzberger, J. 2007, Solar Phys., in press

\bibitem[Wiik et al.(1999)]{ph-wiik99}
Wiik, J.E., Dammasch, I.E., Schmieder, B., \& Wilhelm, K. 1999, Solar Phys., 187, 405

\bibitem[Yakovkin \& Zel'dina(1975)]{ph-yakzel75}
Yakovkin, N.A., \& Zel'dina, M. Yu. 1975, Solar Phys. 45, 319

\bibitem[Zharkova(1989)]{ph-zm89}
Zharkova, V.V. 1989, in Proc. IAU Coll. 117, Hvar Obs. Bull., 13, 331

\bibitem[Zirker \& Koutchmy(1990)]{ph-zirkou90}
Zirker, J.B., \& Koutchmy, S. 1990, Solar Phys., 127, 109 

\bibitem[Zirker \& Koutchmy(1991)]{ph-zirkou91}
Zirker, J.B., \& Koutchmy, S. 1991, Solar Phys., 131, 107

\bibitem[Zirker et al.(1998)]{ph-zirker98}
Zirker, J.B., Engvold, O., \& Martin, S.F. 1998, Nature, 396, 440

\end{thebibliography}

\end{document}